# High-Resolution Pelvic MRI Reconstruction Using a Generative Adversarial Network with Attention and Cyclic Loss

Guangyuan Li[a], Jun Lv[a,*], Xiangrong Tong[a], Chengyan Wang[b,*], Guang Yang[c,d]

a. School of Computer and Control Engineering, Yantai University, Yantai, China;
b. Human Phenome Institute, Fudan University, Shanghai, China;
c. Cardiovascular Research Centre, Royal Brompton Hospital, SW3 6NP, London, U.K.;
d. National Heart and Lung Institute, Imperial College London, London, SW7 2AZ, U.K.;

**ABSTRACT**

Magnetic resonance imaging (MRI) is an important medical imaging modality, but its acquisition speed is quite slow due to the physiological limitations. Recently, super-resolution methods have shown excellent performance in accelerating MRI. In some circumstances, it is difficult to obtain high-resolution images even with prolonged scan time. Therefore, we proposed a novel super-resolution method that uses a generative adversarial network (GAN) with cyclic loss and attention mechanism to generate high-resolution MR images from low-resolution MR images by a factor of 2×. We implemented our model on pelvic images from healthy subjects as training and validation data, while those data from patients were used for testing. The MR dataset was obtained using different imaging sequences, including T2, T2W SPAIR, and mDIXON-W. Four methods, i.e., BICUBIC, SRCNN, SRGAN, and EDSR were used for comparison. Structural similarity, peak signal to noise ratio, root mean square error, and variance inflation factor were used as calculation indicators to evaluate the performances of the proposed method. Various experimental results showed that our method can better restore the details of the high-resolution MR image as compared to the other methods. In addition, the reconstructed high-resolution MR image can provide better lesion textures in the tumor patients, which is promising to be used in clinical diagnosis.

## 1 INTRODUCTION

Magnetic resonance imaging (MRI) has the characteristics of non-invasive, non-radiation, and high contrast, etc. It is an important medical imaging modality. However, its imaging speed is relatively slow, which is mainly limited by the physiological factors including magnetic field strength, slew rate, nerve stimulation, etc. \cite{lustig2008compressed} {donoho2006compressed}. Therefore, the slow imaging speed has become the main factor affecting the development of MRI technology. The slower imaging speed will cause long-term data collection and cause discomfort to the patient. Severe motion artifacts can also be introduced from the patient's movements. Therefore, improving the imaging speed of MRI has great clinical and economic significance. The current research proposes that the method of image reconstruction can be used to accelerate MRI scanning \cite{eo2018kiki} {2020Deep} {Hollingsworth2015Reducing}, and has achieved desired results. These methods used deep learning to reconstruct under-sampled images without artifacts. With the introduction of super-resolution methods, we can also apply super-resolution methods to accelerated MRI imaging. We can get low-resolution (LR) images through fast scanning, and high-resolution (HR) images through trained neural networks, and can also achieve the purpose of accelerating MRI imaging.

The first method of applying deep learning to super-resolution reconstruction was SRCNN \cite{dong2014learning}, and its network structure was composed of three convolutional layers. The proposal of FSRCNN \cite{dong2016accelerating} improved the previous problem of SRCNN that needed to interpolate LR images before they can be used as the input of the network. The FSRCNN can directly input the original LR images. ESPCN \cite{shi2016real} proposed an efficient method to extract features directly on the size of LR images to obtain HR images. VDSR \cite{kim2016accurate} added ResNet to the super-resolution processing method, had a positive impact on the subsequent deep learning super-resolution method. DRCN \cite{kim2016deeply} used the previous RNN structure in super-resolution processing for the first time, and deepened the network structure through residual learning. The RED \cite{mao2016image} used an encoder-decoder framework in the super-resolution method. The network structure was composed of a symmetrical convolutional layer-deconvolutional layer, which can learn an end-to-end mapping from LR images to the corresponding HR images. DRRN \cite{tai2017image} combined the ideas of ResNet, VDSR, and DRCN methods, and proposed a deeper network structure to the super-resolution task. Experiments proved that a deeper network can obtain better results. SRDenseNet \cite{tong2017image} input the features of each layer in the dense block to all subsequent layers to make their features concatenate, reducing the risk of gradient disappearance. SRGAN \cite{ledig2017photo} used the generative confrontation network (GAN) \cite{goodfellow2014generative} in the super-resolution method, using perceptual loss and generative adversarial loss, which effectively improved the effect of the network. EDSR \cite{lim2017enhanced} was an improvement to SRResNet \cite{ledig2017photo}, and further deepening the network structure to obtain deeper details and improved the effectiveness of the network.

Many research methods had shown that super-resolution can not only be used in natural image processing but also can achieve excellent results in MRI reconstruction. Qui et al. \cite{qiu2020super} proposed an effective medical image super-resolution for the reconstruction of the knee images. The network proposed by this method is composed of three hidden layers and a sub-pixel convolutional layer. Lyu et al. \cite{lyu2020mri} applied deep learning and ensemble learning to the super-resolution of MR images. Steeden et al. \cite{steeden2020rapid} used the a novel method to super-resolution reconstruction of the LR whole heart bSSFP data quickly collected in the clinical environment, and used the 3D residual U-net \cite{ronneberger2015u} as the training network, and obtained the expected effect. Zheng et al. \cite{zheng2020hybrid} proposed a hybrid network (HybridNet) to be used in super-resolution MR image reconstruction. Du et al. \cite{du2020super} proposed an anisotropic MR image super-resolution reconstruction method based on convolutional neural network (CNN) \cite{LeCun1989Backpropagation} residual learning with long-skip connections and short-skip connections. Sun et al. \cite{sun2020high} applied the deep convolutional GAN \cite{radford2015unsupervised} to the HR reconstruction of Breast MRI and obtained good reconstruction results. Chaudhari et al. \cite{reference:chaudhari2018super} proposed a method of using CNN to perform knee MR image super-resolution. Lyu et al. \cite{reference:2019 LyuMS} proposed a method using progressive network for multi-contrast super-resolution MRI.

In addition, researches have proven that the use of GAN can effectively reconstruct under-sampled images into images close to clinical medical standards for accelerating MRI. Yang et al. \cite{reference:yang2017dagan} uses deep de-aliasing GAN to perform CS reconstruction. Mardani et al. \cite{reference: gancs2019} proposed a GAN for under-sampling MRI reconstruction, and used least-squares GAN and pixel-wise L1 as the cost function during training. Quan et al. \cite{reference:2018Compressed} proposed the use of cyclic loss in the GAN for end-to-end MRI reconstruction. Lv et al. \cite{reference:lv2020parallel} proposed the use of GAN to remove g-factor artifacts in SENSE reconstruction and proposed the application of a combination of parallel imaging and GAN to multi-channel MRI Image reconstruction \cite{reference:lv2021pic}. Korkmaz et al. \cite{reference:yurt2020semi} proposed a novel unsupervised SLATER for MRI reconstruction, which was composed of a deep adversarial network with cross-attention transformer blocks. Dar et al. \cite{reference: Dar2020} used Conditional GAN to perform synergistic recovery of undersampled multi-contrast acquisitions. The GAN-based method can also be applied to image synthesis. Dar et al. \cite{reference: Dar2019} proposed the use of Conditional GAN for multi-contrast MR image synthesis. This method enhances the synthesis performance of the network by using a pixel-wise loss in registered multi-contrast images and a cyclic loss in unregistered images. Yurt et al. \cite{reference:korkmaz2021unsupervised} proposed a semi-supervised depth generation model for MRI image synthesis, which can generate MR images with the same quality as the fully-supervised model. Shamsolmoali et al. \cite{reference:shamsolmoali2021image} summarized some recent researches on the use of GAN for image synthesis.

It is worth noting that the resolution of MR images is an important factor in clinical medical diagnosis. High-resolution MR images can provide radiologists with more information and facilitate future analysis. However, it takes a longer time to obtain a higher resolution MR image. In some circumstances,, even if the scan time is extended, it is still difficult to obtain HR images. Therefore, the use of super-resolution methods to transform LR MR images into HR MR images through deep learning is helpful for diagnosis in clinical medicine. Based on the above research methods, we proposed a novel super-resolution reconstruction method that uses the attention mechanism and cyclic loss in a GAN. In the following sections, we will introduce our methods, experiment settings, experiment results, discussion, and conclusion in detail.

A summary of our main contributions in this study are listed below:

(1) We proposed a novel GAN-based super-resolution reconstruction method, and used the pelvic dataset in the experiment;

(2) We proved that the use of cyclic loss and attention in GAN can be well applied to super-resolution reconstruction;

(3) We analyzed the performance of each super-resolution method under different upsampling factors.

## 2  METHOD

### 2.1 low-resolutio (LR) Data

We designed a simulation process to obtain LR MR images. We used a LR mask $L_m$, sampling from the center in the horizontal direction. The sampling rate is 50% and 25%, which represent downsampling 2× and 4×, respectively. We used Fourier Transform (FT) to transform the original image $HR_{gt}$ into k-space image $HR_k$, and then used Lmask to remove part of the high-frequency information in $HR_k$ to obtain $L_k$. Then through the inverse Fourier transform (IFT) to transform L$_k$ into the image domain space, and the LR image of the same size as HR$_{gt}$ can be obtained as the input of the network. The process can be expressed as:

$$LR = \bar{\mathcal{F}}(L_k) = \bar{\mathcal{F}}\big(\mathcal{F}(HR_{gt}) \odot L_m\big) \quad (1)$$

where $\mathcal{F}$ stands for FT, $\bar{\mathcal{F}}$ stands for IFT, $\odot$ means dot multiplication operation.

## 2.2 Generative Adversarial Network

Many studies demonstrated the excellent effect of GAN for accelerating MRI scanning \cite{lv2021pic} {yang2017dagan} {arjovsky2017wasserstein} {zhu2017unpaired} {guo2020deep}. The super-resolution method we proposed is also based on GAN, which has strong generation and learning capabilities. GAN is implicitly used as the optimization target through the discriminator. GAN is composed of generator G and discriminator D. The generator learns and extracts details of the input real image *Real*. Its purpose is to generate a fake image *Fake* through the learned information to let D distinguish. This is exactly the work of D, to determine whether the input image is real image *Real* or image *Fake* generated by G. G can continuously learn the effective information in the real image *Real* to continuously generate *Fake* image. In order to correctly distinguish *Real*/*Fake*, D also continuously optimizes in the process. In the continuous learning and optimization of G and D, the generative adversarial loss function $L_{gan}$ is generated. GAN continuously optimizes the network through $L_{gan}$, so that the image generated by G can be infinitely close to the real image. In our proposed network, $L_{gan}$ can be expressed as:

$$\min_G \max_D L_{gan}(D,G) \quad (2)$$

where

$$L_{gan}(D,G) =$$

$$E_{z \sim P_G(z)}\big[1 - logD(G(LR))\big] + E_{x \sim P_{data}(x)}\big[logD(HR_{gt})\big] \quad (3)$$

where $E(*)$ represents the expected value of the distribution function, $P_G(z)$ represents the distribution of data learned by the generator, and $P_{data}(x)$ represents the distribution of data in the real MR image training set. $D(HR_{gt})$ means to discriminate the real *HR* image, and we expect its discrimination result to be as close to *1* as possible. $G(LR)$ represents the *HR* image generated by the generator, namely *HR$_{re}$*. For the generated image, we expect the discriminant result $logD(G(LR))$ of the discriminator to be as close to *0* as possible, which is *maxD*. For the optimization of G, it is only necessary to make the discrimination result $logD(G(LR))$ close to *1*, which is *minG*. In this adversarial process to find the global optimal solution and make G achieve the best generation effect.

### 2.2.1 Generator Architecture

In our research, the generator architecture consists of two residual u-nets with encoder and decoder \cite{2017Compressed}. As shown in Figure 1(a)(b), each residual u-net consists of four encoder blocks and four decoder blocks. There is a full-attention behind each decoder block to further extract the details of the feature maps. There is a global long-skip connection between the encoder and the corresponding decoder, which is used to deepen the network level in order to extract the details in the image. The orange part represents the encoder block, the green part represents the decoder block, and the red part represents full-attention. The number of feature maps for each encoder and decoder block is shown in Figure 1(a). In addition, the input of each module is a 4D tensor, and 2D convolution is used for convolution operation, the filter size is 3×3, the stride is 2. The encoder block adopts the down-sampling method, and each encoder block is composed of two residual blocks with short-skip connections (SR-block). SR-block has three layers: 1) The first layer is 2D convolution with a stride of 2, as shown in the blue part of Figure 1(b). 2) The second layer is a residual block with short-skip connections, including conv0, conv1, conv2, and full-attention. Full-attention will be introduced in the following section. In the residual block, conv0 and conv2 are 2D convolutions with a filter size of 3×3 and stride of 1, and the number of feature maps is 64. Conv1 is a 2D convolution with a stride of 1, the filter size is 3×3, and the number of feature maps is 32, as shown in the cyan part of Figure 1(b). 3) The third layer is 2D convolution with a stride of 1, which is used to adjust the feature maps that obtain attention activation. Up-sampling is used for the decoder blocks, and each decoder block is also composed of two SR-blocks. The difference between it and the encoder block is that the first layer of the SR-block is 2D convolution with a stride of 1, and the third layer is 2D convolution with a stride of 2. After the last decoder block, 2D convolution is used to obtain the global information of the feature maps, with a filter size of 3×3 and stride of 1. During the convolution operation, the same mode is used to obtain an output image of the same size as the original image. In addition, studies have confirmed \cite{2018Compressed} using two residual u-nets can extract more image details than using a single residual u-net to generate an image closer to $HR_{gt}$.

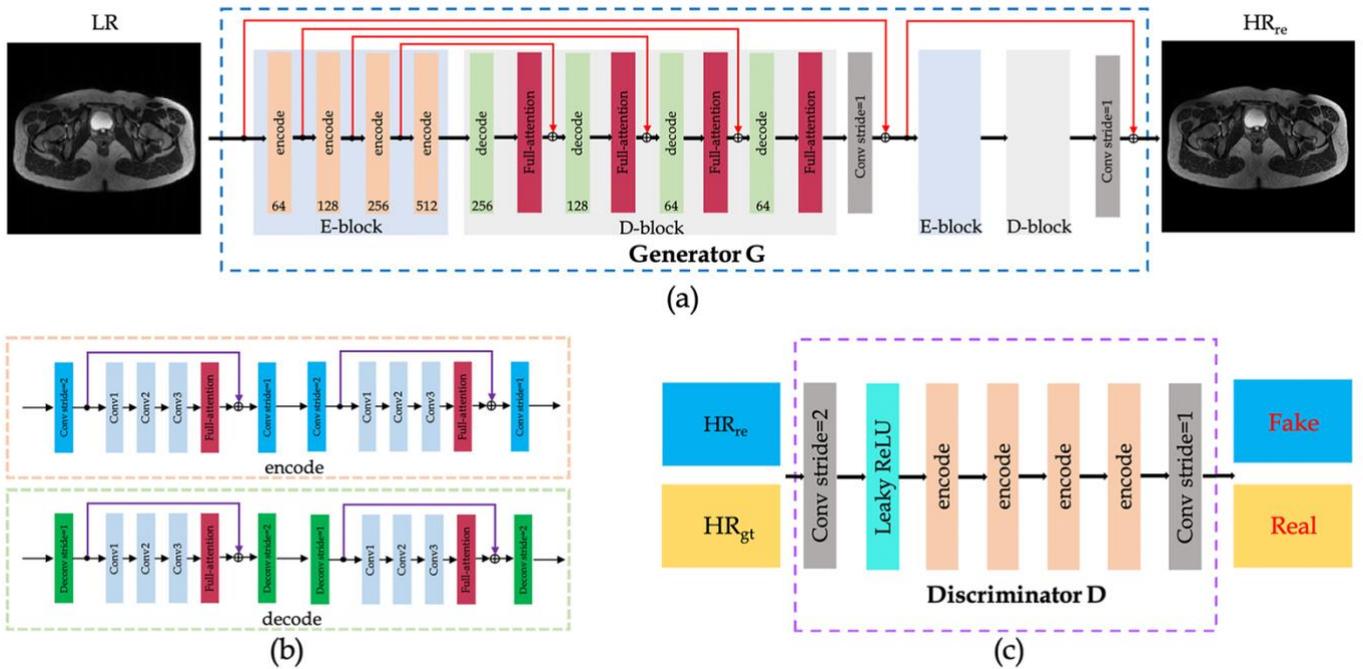

Figure 1. (a) The generator network architecture proposed by our research. The generator takes LR images as input to generate $HR_{re}$ images. The red arrow is global long shortcut. The red block is full-attention. (b) The architecture of encode and decode, in which full-attention is used in each residual block. The purple arrow is local short shortcut. (c) The architecture of the discriminator, D discriminates the generated image $HR_{re}$ and the real image $HR_{gt}$, and the result is Real/Fake.

**2.2.2 Discriminator Architecture**

The architecture of the discriminator we used is composed of a 7-layer network sequential cascade, as shown in Figure 1(c). The first layer is 2D convolution with a filter size of 4×4, a stride of 2, and the number of feature maps is 64. The second layer uses the Leaky ReLU activation function to activate. The third to sixth layers use the same network structure as the encoder in G. The seventh layer uses 2D convolution with a stride of 1 to obtain global information. In addition, we used the output of the sixth-level residual block to evaluate the generative adversarial loss $L_{gan}$. There are two evaluation results of D: one is the real *HR* image $HR_{gt}$, and D should give the result *True*. The other is the *HR* image $HR_{re}$ generated by G. The result of D should be *False*.

## 2.3 Cyclic loss

In this study, we used cyclic consistency loss as an additional constraint when training the network. With enough datasets, any HR image *HR$_{re}$* generated by G has a complete mapping with the low-resolution image *LR*. But this is only an ideal situation. In reality, when the dataset is limited, the cyclic consistency loss can be used to correctly map *LR[m]* and *HR$_{re}$[m]*, where *m* is the entire mapping of the data. The cyclic loss *L$_{cyc}$* in our method is divided into the cyclic loss *L$_{fre}$* in the frequency domain and the cyclic loss in the image domain *L$_{img}$*.

1) The frequency domain cycle loss is that in the *i*-th cycle, the *HR$_{re}$[i]* generated by G is transformed into the frequency domain by the FT to obtain *KS$_{re}$[i]*. Then minimize the difference between *KS$_{re}$[i]* and *KS$_{lr}$[i]*. Next, *KS$_{lr}$[i]* is converted to *LR[i]* as the input of G through the IFT. Among them, *i* is used to marking each cycle, and the input of the *i*-th G is *LR[i-1]*. Therefore, *L$_{fre}$* can be expressed as:

$$L_{fre}(G) = \mathcal{M}(KS_{lr}[i], KS_{re}[i]) \quad (4)$$

where $\mathcal{M}$ represents the mean absolute error.

2) Image domain loss is to ensure that *HR$_{gt}$[i]* and *HR$_{re}$[i]* that generated by G in the *i*-th cycle are as similar as possible. In the *i*-th cycle, *HR$_{gt}$[i]* is transformed by FT and IFT to obtain *LR[i]*, and next *HR$_{re}$[i]* is generated by G. *L$_{img}$* is expressed as:

$$L_{img}(G) = \mathcal{M}(HR_{gt}[j], HR_{re}[j]) \quad (5)$$

In addition, $L_{fre}(G)$ and $L_{img}(G)$ only restrict G, and do not affect D. The loss function of our proposed network is composed of generative adversarial loss (2)(3) and cyclic loss (4) (5), so it can be expressed as:

$$L_{HPMR} = L_{adv}(G, D) + \alpha L_{fre}(G) + \beta L_{img}(G) \quad (6)$$

where $\alpha$ and $\beta$ are hyperparameters used to constrain $L_{fre}(G)$ and $L_{img}(G)$ to stabilize the network during training. We set $\alpha$=1.0 and $\beta$=10.0.

## 2.4 Full-attention

We used the attention mechanism to extract the details in the MR image. The network can learn the attention mechanism autonomously. By assigning different attention weights, the neural network can focus on different key points in the image. There had been studies using the attention mechanism in MR image reconstruction and had shown excellent results \cite{huang2019mri} {2019Self} {yuan2020sara} {huang2020cagan}. In our proposed method, full-attention refers to all-round attention, attention in spatial and channel-wise, as shown in Figure 2. When the feature maps are input to the full-attention module, the spatial weight of the feature map is obtained through spatial attention, and then the weight of each channel is obtained through channel-wise attention. Full-attention can enable feature maps to obtain the distribution of attention weights in both breadth and depth, and extract deep-level image details.

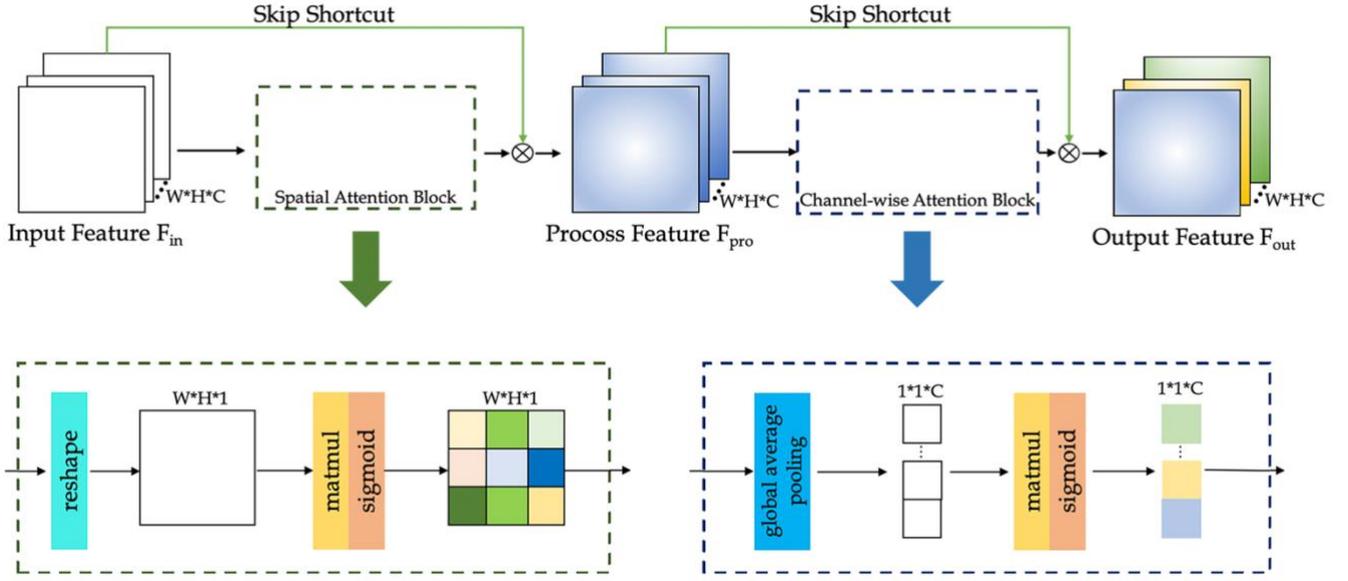

Figure 2. Full-attention architecture, including spatial attention module and channel-wise attention module.

**2.4.1 Spatial attention**

The feature maps $F_{in}$ (shape W×H×C) is input into the spatial attention module (SAB), as shown in Figure 2 (green dashed box). Through the reshape operation, the input $F_{in}$ is calculated at the same position to mean, that is, the number of channels is compressed to 1, and the feature map of W×H×1 is obtained, named $F_s[i]$, (1≤i≤m), where m is the size of the entire feature map, (m=W×H), which means different areas in the feature map. We define a random weight $W_s[i]$ in the matmul operation to assign spatial attention weights to $F_s[i]$. After passing the sigmoid activation function, $F_s$ obtains the distribution of spatial attention weights. Different colors in Fig.1(b) represent different weights. Then according to the idea of residual learning \cite{he2016deep}, we use a short-skip connection with $F_{in}$ for fine-tuning. Therefore, feature maps with spatial attention are obtained, which we call process feature $F_{pro}$, and the shape is W×H×C. The spatial attention is expressed as:

$$F_{pro} = \Phi(\Theta(W_s[i], F_s[i]) + b_s) \otimes F_{in} \quad (7)$$

where $\Theta$ represents the multiplication of matrix $W_s[i]$ and matrix $F_s[i]$, $\otimes$ represents the multiplication of corresponding elements in the matrix, $\Phi$ represents the sigmoid activation function, and $b_s$ is a biases term we set.

**2.4.1 Channel-wise attention**

We assign the channel-wise attention weight to the $F_{pro}$ that has obtained the spatial attention, as shown in the blue dashed box in Figure 2. We use global average pooling to extract feature maps with spatial information into feature map $F_c[i]$ on each channel, with the shape of 1×1×C, 1≤i≤c, where c represents the total number of channels. Like the way of spatial attention distribution, the distribution of channel attention is completed by random weight $W_c[i]$ and sigmoid activation function, so that each channel gets different weights, as shown in different colors in Fig.1(b). Finally, through residual learning with $F_{pro}$ for detail correction, the full-attention activated feature maps $F_{out}$ are obtained. Full-attention can be expressed as:

$$F_{out} = \Phi(\Theta(W_c[i], F_c[i]) + b_c) \otimes F_{pro} \quad (8)$$

Using full-attention in the decoder process can ensure that the detailed information in the image is better restored during the up-sampling process.

## 3 EXPERIMENT

### 3.1 Pelvic Datasets and Model Training

As mentioned in the previous section, The pelvic dataset we used is composed of healthy subjects and patients, and they were all obtained through different MR sequences, which were T2, T2W SPAIR, and mDIXON-W. Acquisition parameters of the pelvic dataset are shown in Table 1. In the pelvic dataset of healthy subjects, a total of 80 people's pelvic MR data participated in the training and testing of the network, of which 64 people were used for network training and 16 people were used for network testing. In addition, we took the pelvic MR dataset of 7 patients for network testing. In the T2 dataset, we took the middle 100 slices of each subject as the experimental data. In addition, we resized the original MR image obtained by scanning to 512×512 as the HR label.

Table 1. Acquisition parameters of the pelvic dataset.

| Dataset | Contrast | MR sequences | Scanning direction | Parameters | Scan time |
|---|---|---|---|---|---|
| T2 | T2W | TSE-3D | TRA | TR: 1250ms; TE: 130ms; Slices: 200; Slice gap: -1mm; FOV: 400mm; Reconstruction matrix: 1120 | 2min30s |
| T2W SPAIR | T2W | TSE-SPAIR | SAG | TR: 4000ms; TE: 70ms; Slices: 24; Slice gap: 1mm; FOV: 250mm; Reconstruction matrix: 528 | 2min16s |
| mDIXON-W | T1W | TSE-mDIXONW | TRA | TR: shortest; TE: shortest; Slices: 80; Slice gap: -2mm; FOV: 250mm; Reconstruction matrix: 512 | 1min04s |

We used Tensorpack \cite{tensorpack}, a neural network training visualization tool based on TensorFlow \cite{tensorflow}, for network visualization training. When training and testing the network, the graphics processing unit (GPU) used is NVIDIA Tesla V100 (4×16GB). The optimizer used by our network was the ADAM algorithm \cite{sun2016deep} {yang2017admm}, the learning rate was $1e^{-4}$, the epoch was 500, and the batch size was 100.

### 3.2 Assessment Method

We used Bicubic, SRCNN, SRGAN, EDSR as the comparison method of our method HPMR. We upsampled the healthy subjects pelvic LR datasets and the patients pelvic LR datasets by 2× and 4× respectively for testing. In order to evaluate the difference between our method and each comparison method, we used peak signal to noise ratio (PSNR), Structural similarity (SSIM), root mean square error (RMSE), and variance inflation factor (VIF) to evaluate the image quality of the HR generated by each method. In addition, we also analyzed the histogram and KL divergence of the result images.

## 4 RESULT

## 4.1 Healthy Subjects

We used the pelvic dataset of healthy subjects to test with an upsampling factor of 2×. The results of each method were shown in Figure 3. We selected slices with similar layers in the three data to display the results. Intuitively, the LR image cannot clearly see the details in the pelvic. Although the Bicubic method effectively improved the LR image, it still needed to be improved intuitively. SRCNN and other comparison methods can effectively reconstruct HR images, but in some details, as indicated by the red arrow in the figure, it needed to be improved compared to HR. The method we proposed, the HR image reconstructed by HPMR, not only had a great improvement in the senses, but also restored the details of the original HR image. In addition, the generated HR image can better distinguish the uterine and the surrounding parts.

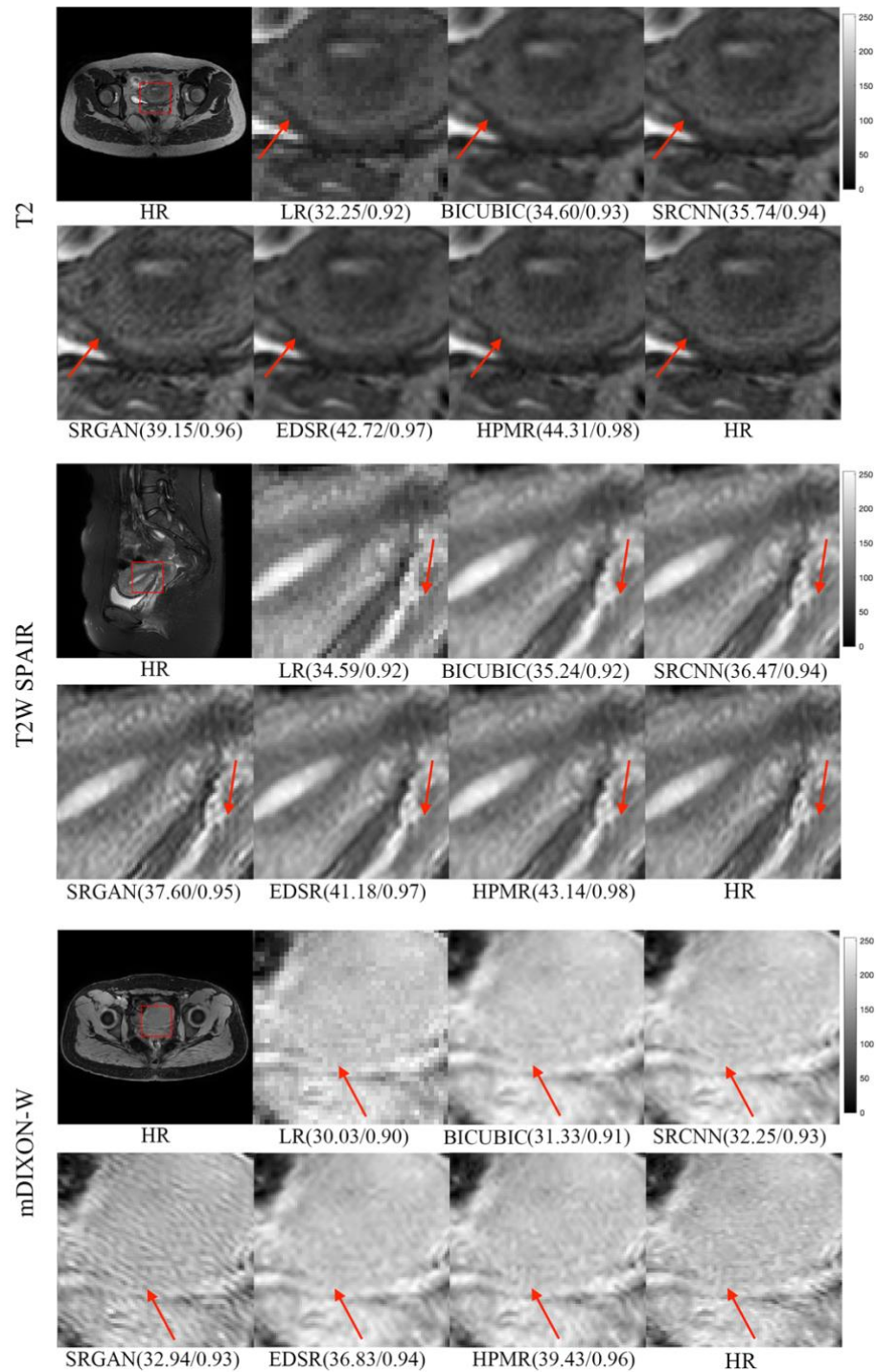

Figure 3. The results of each method when the upsampling factor is 2×. The figure shows the three datasets under healthy subjects pelvic, and the names of the corresponding datasets have been marked in the figure. Among them, the index value (PSNR/SSIM) of each method is marked in the figures.

Figure 4 showed the HR image reconstructed after upsampling 4× the pelvic LR dataset of healthy subjects. It can be seen from the crop images of each method that the quality of the LR image was poor, and the details in the pelvic cannot be distinguished normally. Although the Bicubic method can effectively recover the information lost in the LR, it still cannot get the desired effect from the sensory point of view. Compared with our proposed method, contrast methods such as EDSR had slightly worse recovery at the edge of the uterus, as shown by the red arrow in the figure. In addition, compared with the original HR, HPMR recovered the detailed information in the image as much as possible, and obtained the desired effect.

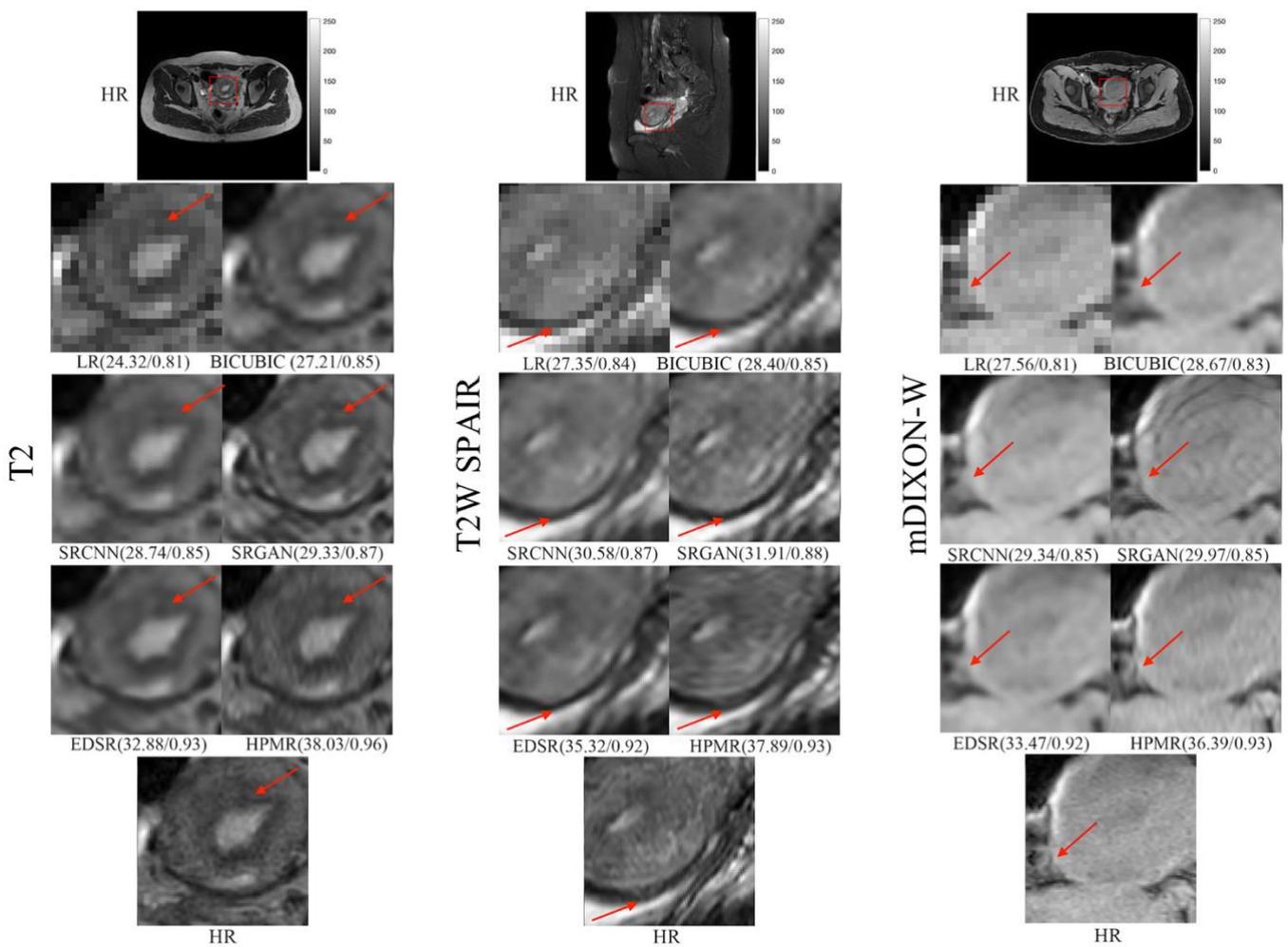

Figure 4. The results of each method when the upsampling factor is 4×. The first column is the T2 data; the second column is the T2W SPAIR data; the third column is the mDIXON-W data. The cropped part in $HR_{re}$ reconstructed by each method intuitively shows the reconstruction effect of each method. It can be seen from the index values of PSNR and SSIM that our proposed method is better than other methods.

We also calculated the KL divergence of each result in Figure 3 and Figure 4 to show the similarity of the distribution of HR images reconstructed by each method, as shown in Table 2. It can be seen that the LR and BICUBIC methods were quite different from the original HR. Deep learning-based methods such as SRCNN, SRGAN, and EDSR had a smaller KL divergence value, which was similar to the original HR. Among them, our proposed method HPMR was the most similar to the original HR. In addition, we had counted the average values of the indicators of each method when the upsampling factors were 2× and 4× in the three datasets of healthy subjects pelvic, including PSNR, SSIM, RMSE, and VIF, as shown in Figure 5. It can be seen from the figure that the various index values of our method HPMR were better than other comparison methods.

Table 2. KL divergence value of each image in Figure 3 and Figure 4.

| Upsampling factor | Dataset | LR | BICUBIC | SRCNN | SRGAN | EDSR | HPMR |
|---|---|---|---|---|---|---|---|
| 2× | T2 | 0.0333 | 0.0282 | 0.0142 | 0.0126 | 0.0098 | **0.0024** |
| | T2W SPAIR | 0.0245 | 0.0233 | 0.0119 | 0.0084 | 0.0063 | **0.0028** |
| | mDIXON-W | 0.0948 | 0.0879 | 0.0365 | 0.0251 | 0.0127 | **0.0067** |
| 4× | T2 | 0.1764 | 0.1620 | 0.0699 | 0.0585 | 0.0189 | **0.0066** |
| | T2W SPAIR | 0.0521 | 0.0473 | 0.0256 | 0.0217 | 0.0109 | **0.0058** |
| | mDIXON-W | 0.2157 | 0.2043 | 0.0901 | 0.0797 | 0.0350 | **0.0129** |

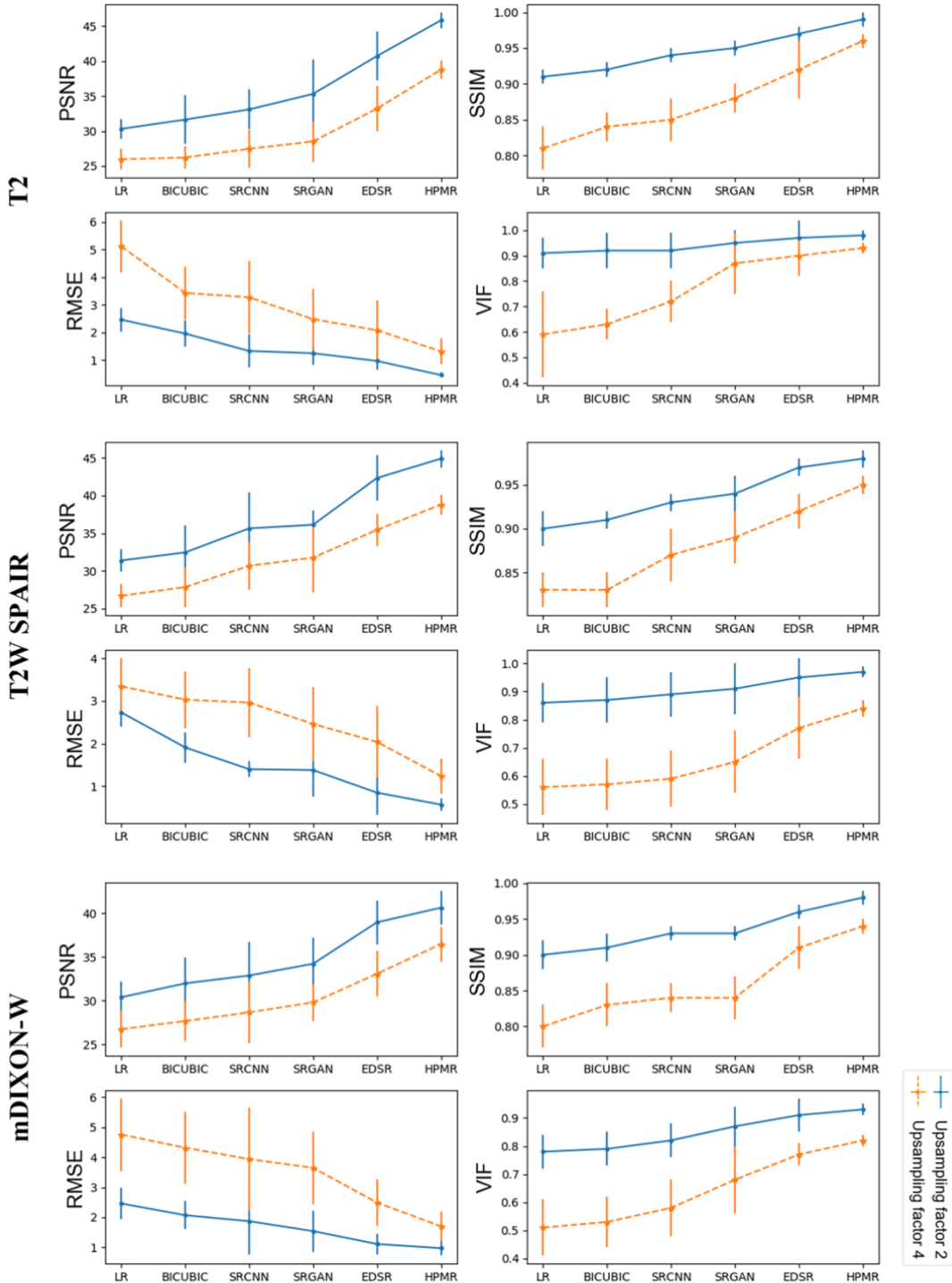

Figure 5. The mean value (std) of each index of the three healthy subjects pelvic dataset. The blue solid line indicates that the LR image is upsampled by 2×, and the orange dashed line indicates that the LR image is upsampled by 4×. PSNR: peak signal to noise ratio; SSIM: structural similarity; RMSE: root mean square error; VIF: variance inflation factor. The unit of RMSE is $10^{-2}$.

### 4.2 Patients

In order to verify the performance of the trained network in the patients dataset, we tested the reconstruction effect of each method on three patients pelvic data with upsampling factors of 2× and 4×. The reconstructed HR image was shown in Figure 6. In Figure 6(a), the LR image can only see the contours of the uterus and uterine fibroids, but cannot clearly see the details. The BICUBIC method can only recover part of the detailed information in the original HR. SRCNN, SRGAN and EDSR can effectively restore the sharpness of the image, but the recovery of the details of the uterine fibroids still needs to be improved. Our proposed method HPMR can better restore the details in the original HR, as shown by the red arrow in Figure 6(a). In addition, we can clearly see the junction of the uterus and uterine fibroids and the internal information of uterine fibroids, which is helpful for radiologists to diagnose.

Figure 6(b) shows the results of each method under the patients pelvic dataset when the upsampling factor is 4×. We can see that the LR image has a low quality and cannot effectively distinguish the uterine part and the uterine fibroids. Although BICUBIC and SRCNN have improved the quality of part of the image, they still cannot effectively distinguish the tumor part in the image from the sensory point of view. Although the SRAGN and EDSR methods are better than the previous comparison methods, the image quality still does not reach the desired effect, and the restoration of the details of the tumor needs to be improved. The HPMR method we proposed, from the junction of the uterus and uterine fibroids, such as the part pointed by the red arrow in Figure 6(b), can effectively distinguish the junction, which was helpful for the doctor to diagnose the size of the patient's uterine fibroids and period. In addition, it can be seen from the image index values (PSNR/SSIM) marked in Figure 6(b) that the quality of the HR image reconstructed by HPMR is higher than other comparison methods.

Table 3. KL divergence value of each image in Figure 6.

| Upsampling factor | Dataset | LR | BICUBIC | SRCNN | SRGAN | EDSR | HPMR |
|---|---|---|---|---|---|---|---|
| 2× | T2 | 0.0274 | 0.0227 | 0.0201 | 0.0193 | 0.0151 | **0.0030** |
|  | T2W SPAIR | 0.0172 | 0.0167 | 0.0141 | 0.0093 | 0.0073 | **0.0047** |
|  | mDIXON-W | 0.0645 | 0.0579 | 0.0404 | 0.0244 | 0.0141 | **0.0089** |
| 4× | T2 | 0.1358 | 0.1246 | 0.0648 | 0.0613 | 0.0541 | **0.0068** |
|  | T2W SPAIR | 0.0747 | 0.0676 | 0.0303 | 0.0344 | 0.0156 | **0.0076** |
|  | mDIXON-W | 0.2341 | 0.2226 | 0.0834 | 0.0811 | 0.0315 | **0.0153** |

We also calculated the KL divergence value of each result image in Figure 6, and the results are shown in Table 3. We can see that the HR image generated by HPMR is most similar to the original HR image. In addition, we also analyzed the histogram of each result in Figure 6(a) when the upsampling factor is 2×. Figure 7 shows the histogram of each image in the T2 dataset, the histogram and quantitative indicators show that HPMR (kurtosis:5.38, skewness:1.35, mean:0.20, Standard:0.09) achieved the closest quantifications to the original HR(kurtosis:5.38, skewness:1.35, mean:0.20, Standard:0.10) as compared to other methods. Other histogram results are shown in the addenda. In addition, we calculated the average value of the index on the three pelvic dataset of 7 patients. Table 4 shows the mean value of indicators for different dataset under different upsampling factors. We can see that when using the patients dataset for testing, the HPMR method has a better performance than other comparison methods.

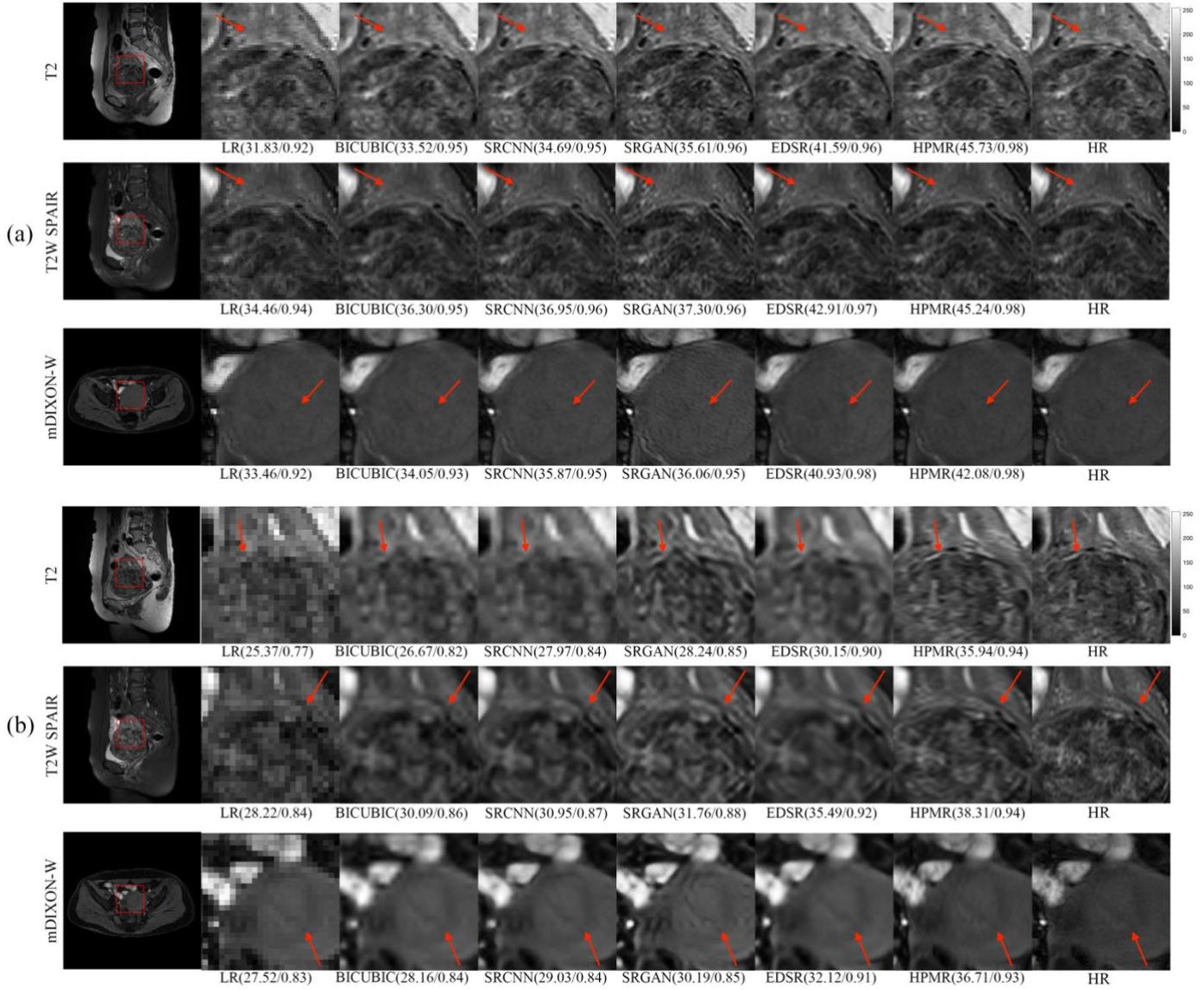

Figure 6. The results of $HR_{re}$ reconstructed by each method when the upsampling factors are 2× and 4× of the patients pelvic data. Among them, (a) is the result when the upsampling factor is 2×, and (b) is the result when the upsampling factor is 4×. The first column of each row is the original HR image, and the rest are the crop images of $HR_{re}$ reconstructed by each method. The red arrow in figure (a) indicates the details of $HR_{re}$; the red arrow in figure (b) indicates the edge of the uterus and uterine fibroids in $HR_{re}$.

### 4.3 Ablation Experiments

In addition, we examined the relative contributions of the various components (cyclic loss, spatial attention, and channel attention) of the HPMR method to network performance. Therefore, we seted up three variant models, 1) using spatial attention and channel attention $HPMR_a$; 2) using cyclic loss and channel attention $HPMR_b$; 3) using cyclic loss and spatial attention $HPMR_c$. Experiments were performed on the T2 pelvic dataset of healthy subjects with an upsampling factor of 4×. Table 5 lists the mean (standard deviation) of each indicator. We can see that the contribution provided by the cyclic loss is the largest, and the spatial attention and channel attention have close contributions.

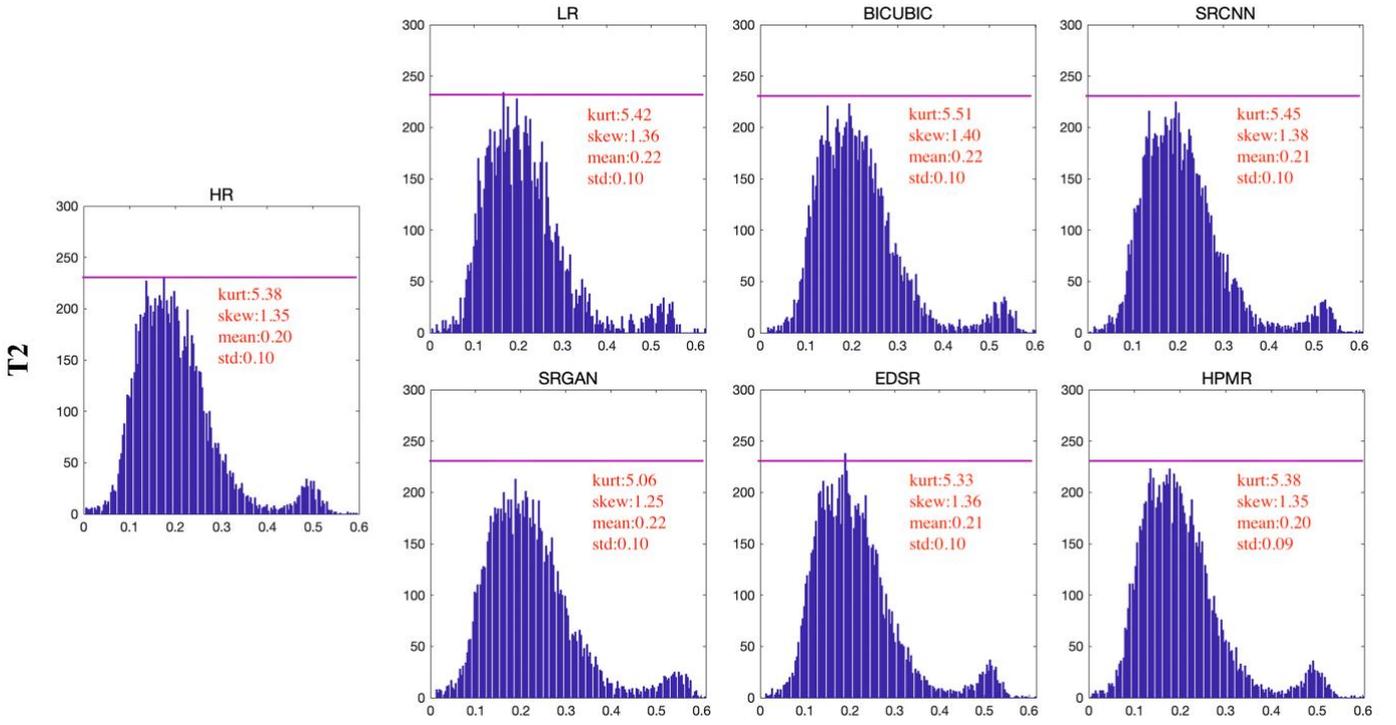

Figure 7. When the upsampling factor is 2×, the histogram of the HR image generated by each method under the T2 patients dataset in Figure 6(a). Kurt: kurtosis; skew: skewness.

Table 4. Means (standard deviations) of each index in the patients pelvic dataset. UP: Upsampling factor; PSNR: peak signal-to-noise ratio; SSIM: structural similarity; RMSE: root mean square error; VIF: variance expansion coefficient. The unit of RMSE is $10^{-2}$.

| Dataset | UF | Method | PSNR | SSIM | RMSE | VIF |
|---|---|---|---|---|---|---|
| T2 | 2× | LR | 32.14±0.94 | 0.92±0.01 | 2.48±0.28 | 0.90±0.04 |
| | | BICUBIC | 33.48±1.04 | 0.95±0.01 | 1.93±0.26 | 0.91±0.04 |
| | | SRCNN | 34.64±1.01 | 0.96±0.01 | 1.91±0.23 | 0.92±0.04 |
| | | SRGAN | 35.36±1.00 | 0.96±0.01 | 1.83±0.35 | 0.93±0.05 |
| | | EDSR | 40.22±3.41 | 0.97±0.01 | 1.68±0.57 | 0.99±0.04 |
| | | HPMR | **46.15±0.71** | **0.99±0.02** | **0.42±0.15** | **0.99±0.01** |
| | 4× | LR | 25.88±1.11 | 0.79±0.02 | 5.12±0.68 | 0.57±0.06 |
| | | BICUBIC | 26.16±1.21 | 0.83±0.02 | 4.43±0.66 | 0.63±0.07 |
| | | SRCNN | 27.45±1.15 | 0.85±0.02 | 3.65±0.83 | 0.65±0.14 |
| | | SRGAN | 27.88±1.13 | 0.85±0.02 | 3.56±0.80 | 0.66±0.09 |
| | | EDSR | 30.38±3.23 | 0.88±0.04 | 3.32±1.43 | 0.80±0.06 |
| | | HPMR | **36.74±1.12** | **0.95±0.01** | **1.44±0.31** | **0.92±0.02** |
| T2W SPAIR | 2× | LR | 34.12±1.34 | 0.93±0.01 | 1.77±0.28 | 0.86±0.05 |
| | | BICUBIC | 37.20±1.45 | 0.95±0.01 | 1.41±0.25 | 0.87±0.04 |
| | | SRCNN | 37.30±1.35 | 0.96±0.01 | 1.39±0.23 | 0.88±0.04 |
| | | SRGAN | 38.94±1.38 | 0.96±0.01 | 1.13±0.38 | 0.94±0.05 |
| | | EDSR | 41.84±2.17 | 0.97±0.01 | 0.85±0.24 | 0.96±0.04 |
| | | HPMR | **44.53±1.12** | **0.98±0.01** | **0.59±0.10** | **0.96±0.01** |
| | 4× | LR | 28.42±1.42 | 0.84±0.02 | 3.43±0.59 | 0.55±0.07 |
| | | BICUBIC | 30.58±1.54 | 0.86±0.02 | 3.01±0.57 | 0.57±0.08 |
| | | SRCNN | 30.57±1.46 | 0.88±0.02 | 3.16±0.65 | 0.57±0.77 |
| | | SRGAN | 31.68±1.28 | 0.88±0.02 | 2.51±0.59 | 0.65±0.09 |
| | | EDSR | 34.99±2.07 | 0.91±0.02 | 1.92±0.63 | 0.75±0.10 |
| | | HPMR | **38.41±1.32** | **0.94±0.01** | **1.27±0.42** | **0.84±0.05** |
| mDIXON-W | 2× | LR | 32.38±1.88 | 0.91±0.02 | 2.46±0.55 | 0.77±0.06 |
| | | BICUBIC | 33.99±1.93 | 0.93±0.01 | 2.07±0.49 | 0.79±0.06 |
| | | SRCNN | 34.14±1.86 | 0.94±0.01 | 2.03±0.46 | 0.81±0.07 |
| | | SRGAN | 35.98±1.78 | 0.95±0.02 | 1.92±0.68 | 0.91±0.06 |
| | | EDSR | 39.40±2.19 | 0.96±0.01 | 1.13±0.31 | 0.93±0.06 |
| | | HPMR | **40.57±1.91** | **0.97±0.01** | **0.96±0.23** | **0.96±0.02** |
| | 4× | LR | 27.19±2.01 | 0.81±0.03 | 4.49±1.11 | 0.48±0.09 |
| | | BICUBIC | 28.17±2.11 | 0.83±0.03 | 4.04±1.06 | 0.49±0.10 |
| | | SRCNN | 28.96±2.02 | 0.85±0.02 | 4.03±1.10 | 0.50±0.10 |
| | | SRGAN | 29.94±1.97 | 0.86±0.02 | 3.85±1.21 | 0.63±0.13 |
| | | EDSR | 32.76±2.58 | 0.92±0.02 | 2.46±0.73 | 0.75±0.16 |
| | | HPMR | **36.23±1.97** | **0.94±0.01** | **1.70±0.53** | **0.81±0.03** |

Table 5. Reconstruction performance in HPMR ablation experiments. The mean value (standard deviation) of each indicator in the T2 pelvic dataset of healthy subjects with an upsampling factor of 4×. The unit of RMSE is 10-2; + means used component; - means unused component.

|  | Method | | | Metric | | | |
| --- | --- | --- | --- | --- | --- | --- | --- |
|  | Cyclic loss | Spatial attention | Channel attention | PSNR | SSIM | RMSE | VIF |
| $HPMR_a$ | - | + | + | 35.96±1.70 | 0.94±0.02 | 1.80±0.71 | 0.87±0.03 |
| $HPMR_b$ | + | - | + | 37.71±1.31 | 0.95±0.01 | 1.52±0.66 | 0.91±0.01 |
| $HPMR_c$ | + | + | - | 37.66±1.29 | 0.95±0.01 | 1.47±0.47 | 0.90±0.02 |
| $HPMR_f$ | + | + | + | **38.76±1.30** | **0.96±0.01** | **1.32±0.46** | **0.93±0.02** |

## 5 DISCUSSION

The purpose of this research is to reconstruct LR images into HR images and restore the details in the original HR images so that the generated HR images can be used in clinical examinations. The significance of this research is that LR images can be obtained through fast MRI scans, which can shorten the scan time, and then use the trained network to reconstruct the LR images into HR images to achieve the purpose of accelerating the MRI scan. In addition, it is difficult to obtain original HR images for some parts, such as the abdomen, organs, and LR images caused by the patient's breathing movement. Therefore, the LR images obtained can be refined by the method we propose to facilitate doctors' diagnosis. The data used in our proposed method is female pelvic data, including normal and abnormal. Uterine fibroids are a common female disease. We use the network trained from normal data to test the abnormal data, and the reconstructed HR image has the expected quality. From Figure 6, it can be seen that the HR image reconstructed by HPMR can clearly distinguish the uterine part and the uterine fibroids, and restore the detailed information, which is beneficial to the diagnosis of doctors. Our method is expected to be used in clinical medicine. SRGAN uses the GAN for the super-resolution method, and EDSR uses a deeper network in the super-resolution method. Both methods can excellently reconstruct LR images into HR images. Based on these two points, we chose to use GAN, and use two cascaded residual u-net as the generator. Each encoder-decoder block in each residual u-net is composed of small residual blocks connected by short hops to increase the depth of the network. In addition to these two points, our proposed method also uses cyclic loss and attention mechanisms. The cyclic loss can make the network use $KS_{lr}[i]$ and $KS_{re}[i]$ cycle in the frequency domain in the case of limited datasets so that the difference between them is smaller so that the network learns more image details. Through the cycle of $HR_{re}[i]$ and $HR_{gt}[i]$ in the image domain, the $HR_{re}$ generated by G is more and more similar to the real $HR_{gt}$. Cyclic loss can play a role in expanding the datasets. We use cyclic loss in HPMR to establish the mapping between LR and reconstructed $HR_{re}$ as completely as possible. Through this mapping, the network can learn more details. The attention mechanism is to simulate the human senses, so that the system can focus on effective information, and can extract the characteristics of the image. What we use in HPMR is soft attention, which focuses on the characteristics of the spatial or channel, and can calculate the gradient through a neural network. In the previous research \cite{huang2019mri}, the attention mechanism was used in residual u-net for MR image reconstruction, and higher quality pictures were obtained. In our proposed method, full-attention is to assign spatial attention weights and channel-wise attention weights to feature maps so that the network can learn the details in the image more effectively, and then generate high-quality $HR_{re}$. The feasibility of using spatial attention and channel-wise attention simultaneously in neural networks has been confirmed in the \cite{chen2017sca} method. According to the tests in the normal pelvic datasets and the abnormal pelvic datasets, the quality of the $HR_{re}$ reconstructed by HPMR is the best, and when the upsampling factor is 4×, the obtained $HR_{re}$ can restore the details of the original HR to the greatest extent. Therefore, our method is expected to be used in clinical medicine in the future.

Although our proposed method can achieve the expected results, there is still room for improvement. When we train the network, we need the original HR image as a label. We hope that in future work, LR images can be used as labels in between, that is, completely separate from the original HR images when training the network. This problem can use unsupervised learning methods, when training the network, let the network learn effective information in the LR label.

# 6 CONCLUSION

The generative adversarial network with cyclic loss and attention is proposed by us for HR reconstruction of the pelvic datasets, including the healthy subjects pelvic datasets and the patients pelvic datasets. Each dataset is obtained through different MR sequences. This method can upsampling LR images by 2× and 4×, and the reconstructed $HR_{re}$ can reach the quality required by radiologists for diagnosis.

# ACKNOWLEDGEMENT

Funding: This study was funded in part by the National Natural Science Foundation of China (61902338), in part by the National Natural Science Foundation of China (No. 62001120) and the Shanghai Sailing Program (No. 20YF1402400), in part by the European Research Council Innovative Medicines Initiative on Development of Therapeutics and Diagnostics Combatting Coronavirus Infections Award "DRAGON: rapiD and secuRe AI imaging based diaGnosis, stratification, fOllow-up, and preparedness for coronavirus paNdemics" [H2020-JTI-IMI2 101005122] and in part by the AI for Health Imaging Award "CHAIMELEON: Accelerating the Lab to Market Transition of AI Tools for Cancer Management" [H2020-SC1-FA-DTS-2019-1 952172].

# AUTHOR

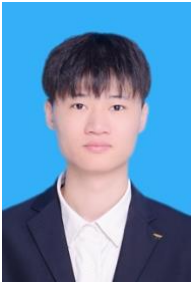

**Guangyuan Li** was born in Jining, Shandong, China in 1997. He is currently pursuing the M.S. degree at the School of Computer and Control Engineering, Yantai University. His research direction is deep learning and medical image reconstruction.

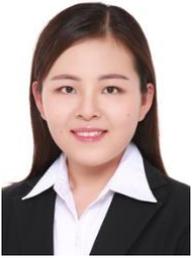

**Jun Lv** was born in Yantai, Shandong, China in 1990. She received the B.S.degree in Intelligence Science and Technology in Xidian University in 2013 and the Ph.D. degree in Biomechanics and Medical Engineering from Peking University in 2018. Her research interests include medical image reconstruction, image registration and image segmentation. She is a Lecturer in the School of Computer and Control Engineering from Yantai University.

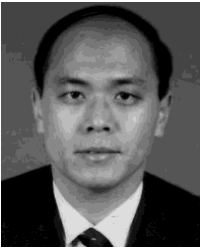

Xiangrong Tong received the Ph.D. degree in computer science and technology from Beijing Jiaotong University. He is currently a Full Professor with Yantai University. He has published more than 50 articles in well-known journals and conferences. He has also presided and joined three national projects and three provincial projects. His research interests include computer science, intelligent information processing, and social networks.

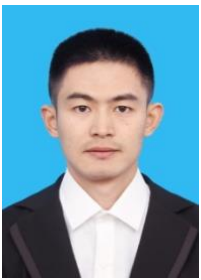

**Chengyan Wang** received the B.S. degree in Biomedical Engineering from Beijing Institute of Technology, Beijing, China, in 2012 and the Ph.D. degree in Biomechanics and Medical Engineering from Peking University, Beijing, China, in 2017.

From 2017 to 2018, he was a Post-doctoral Fellow with the Institute of Biomedical Engineering, Shanghai Jiao Tong University, Shanghai, China. From 2018 to 2019, he was a Post-doctoral Fellow with the Institute of Electronic Engineering, University of Illinois at Urbana-Champaign, IL, USA. Since 2019, he has been an Assistant Professor with the Human Phenome Institute, Fudan University, Shanghai, China. His research interests include image reconstruction, medical image analysis and deep learning.

Dr Guang Yang obtained his M.Sc. in Vision Imaging and Virtual Environments from the Department of Computer Science in 2006 and his Ph.D. on medical image analysis jointly from the CMIC, Department of Computer Science and Medical Physics in 2012 both from University College London. He is currently a tenured senior research fellow at NHLI, Imperial College London and an honorary senior lecturer at the School of Biomedical Engineering & Imaging Sciences, King's College London. He is the head of the Smart Imaging lab funded by UKRI, BHF and ERC. His research interests include pattern recognition, machine learning, and medical image processing and analysis.

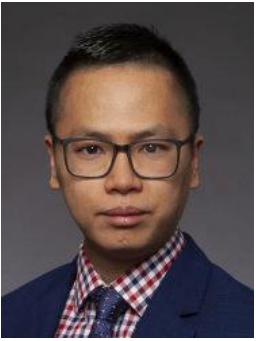

# SUPPLEMENTARY MATERIALS

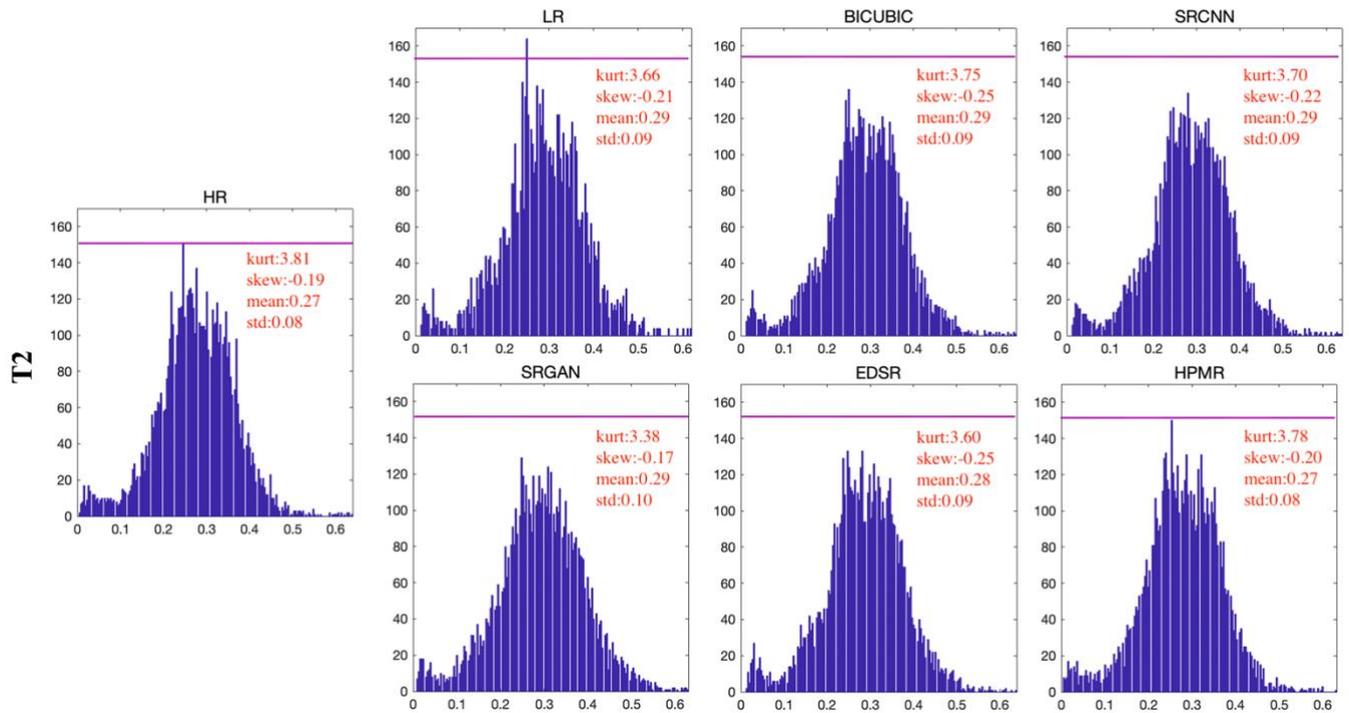

Supp. Figure 1. When the upsampling factor is 2×, the histogram of the HR image generated by each method under the T2 healthy subjects dataset in Figure 3. Kurt: kurtosis; skew: skewness.

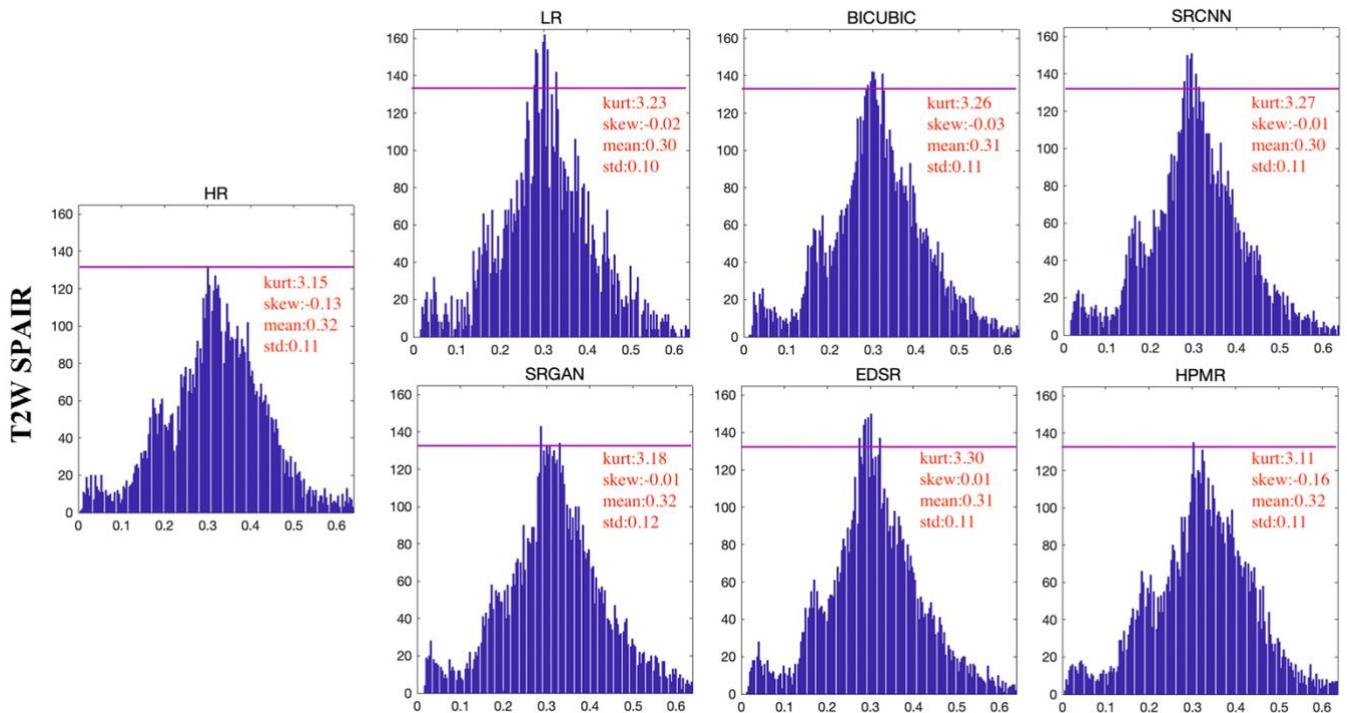

Supp. Figure 2. When the upsampling factor is 2×, the histogram of the HR image generated by each method under the T2W SPAIR healthy subjects dataset in Figure 3. Kurt: kurtosis; skew: skewness.

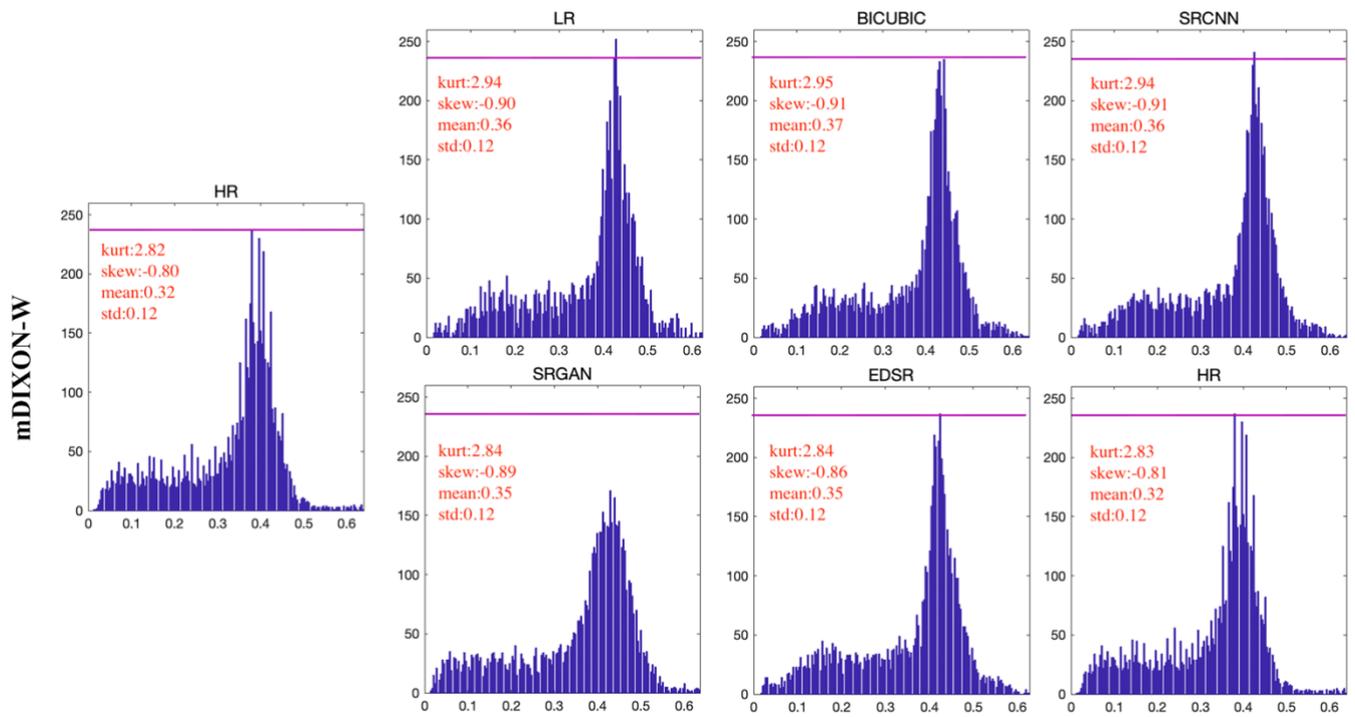

Supp. Figure 3. When the upsampling factor is 2×, the histogram of the HR image generated by each method under the mDIXON-W healthy subjects dataset in Figure 3. Kurt: kurtosis; skew: skewness.

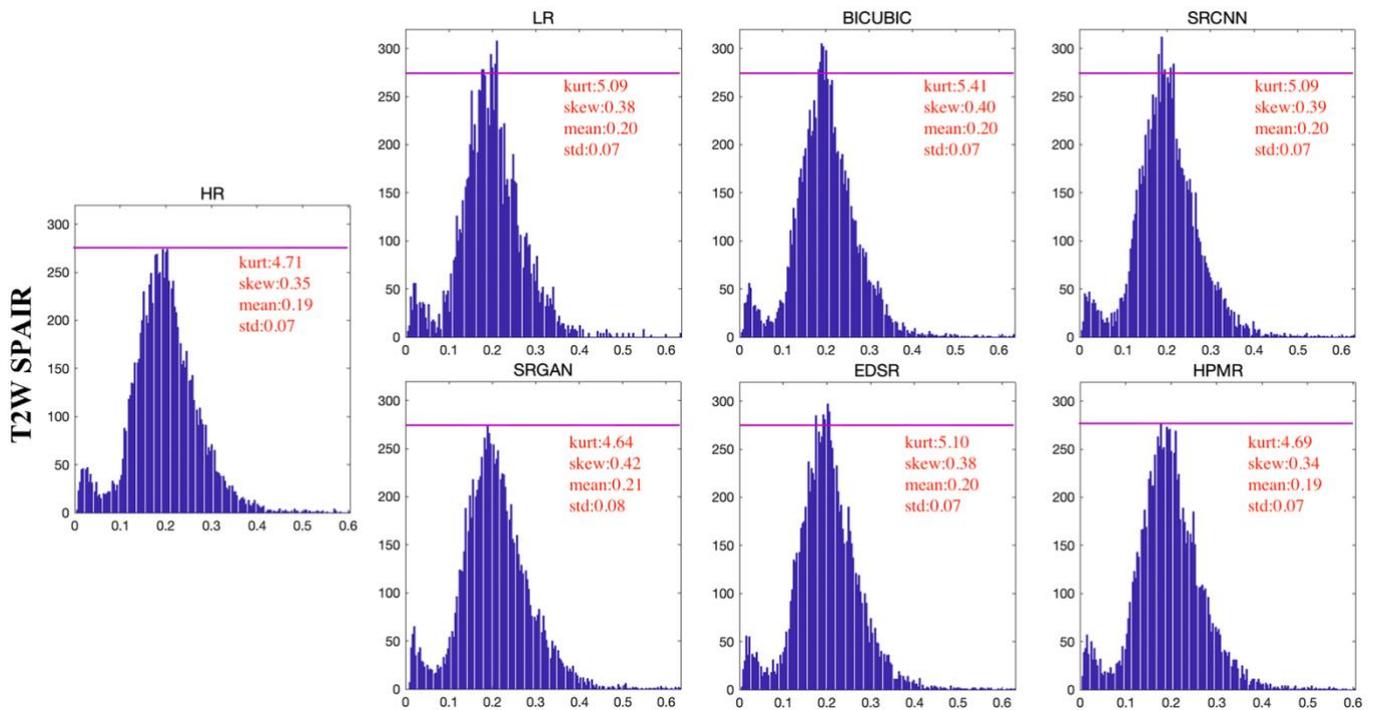

Supp. Figure 4. When the upsampling factor is 2×, the histogram of the HR image generated by each method under the T2W SPAIR patients dataset in Figure 6(a). Kurt: kurtosis; skew: skewness.

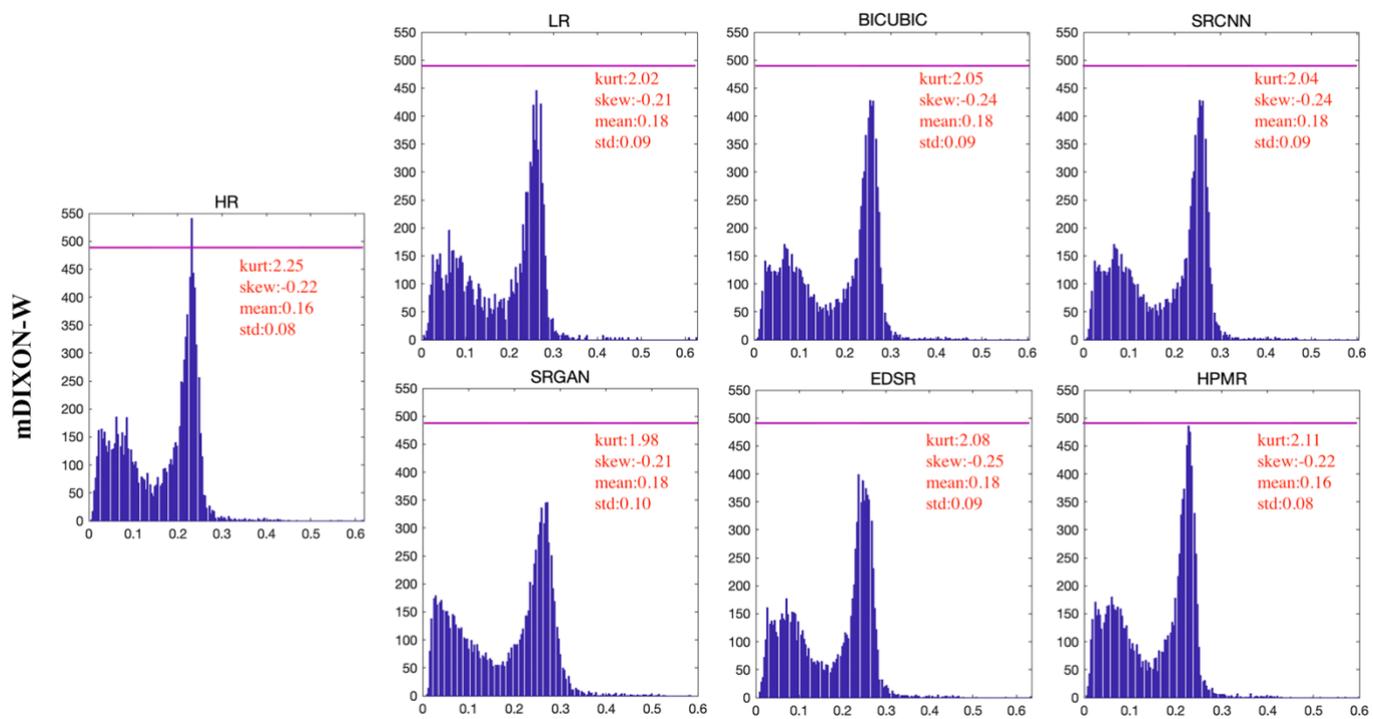

Supp. Figure 5. When the upsampling factor is 2×, the histogram of the HR image generated by each method under the mDIXON-W patients dataset in Figure 6(a). Kurt: kurtosis; skew: skewness.